%% file: RJwrapper-arxiv.tex
\pdfoutput=1
\documentclass[a4paper]{report}
\usepackage[utf8]{inputenc}
\usepackage[T1]{fontenc}
\usepackage{RJournal}
\usepackage{amsmath,amssymb,array}
\usepackage{booktabs}

\usepackage{mathrsfs}
\IfFileExists{xparse.sty}{\usepackage{xparse}}{}

\begin{document}

%% Placeholders kept so the style's internal macros are defined;
%% nothing below is rendered because the head/foot fields are blanked.
\sectionhead{}
\volume{}
\volnumber{}
\year{}
\month{}

\begin{article}
\input{fdwasserstein-working}
\end{article}

\end{document}

%% file: fdwasserstein-working.tex
% !TeX root = RJwrapper.tex
\title{fdWasserstein: Optimal Transport Methods for Covariance Operators of Functional Data}

\author{by Valentina Masarotto}

\maketitle

\abstract{%
Data increasingly arrive as collections of curves --- a voice recording, a growth trajectory, a day of sensor readings --- where each observation is a whole function rather than a single number. The usual question asked of such data is how the average curve differs from one group to the next. But the average is only half the picture: two populations of curves can share almost the same mean and still differ profoundly in how they fluctuate around it, and it is often this variability --- the pattern of covariation within a curve --- that carries the scientific signal. Comparing populations at this level means comparing their covariance operators, and these do not live in a flat space where one can simply subtract or average them. The fdWasserstein package equips R users with functions to make such comparisons. It is centered on the geometry of optimal transport, under which covariance operators can be meaningfully averaged, contrasted, and interpolated. It provides the Procrustes-Wasserstein distance between covariance operators, their Fréchet mean (barycenter), an ANOVA-type permutation test for the equality of several covariances, principal component analysis of covariance variation, and an entropy-regularized soft clustering of curves by their covariance structure. We outline the underlying ideas, discuss the implementation and demonstrate the complete workflow on the phoneme data shipped with the package.
}

\section{Introduction}\label{introduction}

Every day, sensors, clinical wearables, and recording devices collect data that are not simple numbers, but continuous signals
unfolding over time: the pitch of a voice, the motion of a limb, the vibration of a bridge. When data are smooth and
continuous in this way, they are best described as functional and the natural mathematical framework to analyze them is Functional Data Analysis (FDA). FDA is a branch of
statistics that treats each observation not as a vector of numbers, but as a curve or surface in its own right \citep{ramsay2005} or, more technically, as realizations of smooth random processes defined on a continuous domain.
Since the seminal book of \citet{ramsay2005}, FDA has become an indispensable toolkit for the statistical analysis of continuously varying observations.
A key player in
FDA is the covariance operator of a functional process, which encodes the second-order structure of a random process: loosely, it is a mathematical tool that captures not just how much a
collection of curves varies, but how that variation is structured across time: which parts of a curve tend to move together,
and by how much.
It is not a descriptive summary --- it is the foundation for nearly all statistical inference on functional data.
Most notably, it allows one to write the optimal finite-dimensional approximation of a functional curve through the Karhunen--Loève decomposition, which is essential
whenever one needs to work with functional data in practice.
Classical FDA focuses on mean
functions and on the Karhunen--Loève decomposition of a single covariance
operator in order to carry out statistical inference. Yet the mean function tells only half the story.
A recent line of work treats the covariance operators themselves as the
objects of statistical interest: several populations of curves are to be contrasted, summarized or clustered relative to the nature of their dispersion around their means, rather than relative to their means themselves \citep{masarotto2019, masarotto2024, panaretos2010, pigoli2014}. We speak of \emph{second order variation} whenever variation arises in the covariance structure of several groups of functional curves.

Covariance operators do not live in a linear space in any statistically
natural way. Early contributions to second-order inference embedded the operators in the larger, linear space of Hilbert--Schmidt operators; covariance operators, however, form a closed convex cone rather than a linear subspace of that space, and are not closed under linear operations such as differences. Moreover, covariances are
trace-class non-negative operators, so rather than being seen as Hilbert--Schmidt objects themselves, they are better represented as ``squares'' of such operators. \citet{pigoli2014} and \citet{masarotto2019} argue that a
natural metric for covariance operators is the Procrustes distance, and
\citet{masarotto2019} show that it coincides with the 2-Wasserstein distance of
optimal transportation between the centered Gaussian processes having those
covariances (in the matrix case this metric is also known as the Bures
distance; \citet{bhatia2019}). The Wasserstein geometry comes with a rich toolbox:
explicit optimal transport maps, Fréchet means \citep{agueh2011, zemel2019},
and a tangent space structure that linearizes the geometry locally. On this
foundation, \citet{masarotto2024} developed an ANOVA-type test and the first principal
component analysis for collections of covariance operators, and
\citet{masarotto2024clustering} a soft clustering procedure for samples of curves based on
their covariances.

The \CRANpkg{fdWasserstein} package implements this framework in a single, self-contained software environment. Prior to its development, practitioners wishing to perform second-order inference on functional covariance operators were forced to combine disparate scripts, often drawn from supplementary materials of individual papers, with no consistent interface, and no documentation suitable for applied researchers. The package fills this gap, consolidating the methodology of \citet{masarotto2019}, \citet{masarotto2024}, and \citet{masarotto2024clustering} into a coherent, CRAN-hosted toolkit.
Related works are the
\CRANpkg{shapes} package \citep{dryden2009}, which provides distances, including a
Procrustes-type one, and means for symmetric positive semi-definite
matrices, but is oriented towards low-dimensional covariance matrices
arising in shape analysis and diffusion tensor imaging, and offers none of the inferential tools that \pkg{fdWasserstein} includes (testing, PCA, and clustering). Packages such as \CRANpkg{frechet} \citep{frechet-pkg} compute Fréchet means and
regressions for various random objects, and \CRANpkg{fdANOVA} \citep{gorecki2019} tests
equality of mean functions, which is the first-order analogue of the
problem treated here. On the optimal-transport side, \CRANpkg{transport} \citep{transport-pkg} and \CRANpkg{T4transport} \citep{t4transport-pkg} provide general-purpose transport computations between discrete or empirical measures, but neither targets covariance operators nor offers the associated inference.

The package \pkg{fdWasserstein} is available from the Comprehensive R Archive Network at \url{https://CRAN.R-project.org/package=fdWasserstein}.
In this paper we document the package in detail, explain briefly the underlying statistical theory and present worked examples on the built-in phoneme dataset. A full real-data application to phonetics data which demonstrates how the package can be deployed in an end-to-end analysis pipeline alongside established tools such as the \CRANpkg{mgcv} package \citep{wood2017} for generalized additive mixed models can be found in \citet{masarotto2024tonal}.

The remainder of the paper is organized as follows. The next section
recalls the minimal amount of theory required to use the package: the 2-Wasserstein distance, the definition of optimal maps, of Fréchet barycenter and the tangent space. We then describe the transportation-based ANOVA and PCA, and soft clustering. Section 3 describes the package design and the six user-visible functions with their
main implementation choices. The last section walks through a complete
analysis of the phoneme data: testing the equality of the phoneme
covariances, exploring their variation by tangent PCA, and recovering the
phonemes by soft clustering of the raw curves.

\section{The Wasserstein--Procrustes geometry in brief}\label{the-wassersteinprocrustes-geometry-in-brief}

Our data structure assumes \(K\) independent samples \(\{X_{1,j}\}_{j=1}^{n_1}, \dots ,\{X_{K,j}\}_{j=1}^{n_K}\) of i.i.d. random elements in a Hilbert space \(\mathcal{H}\). Each of the \(K\) samples is modeled by a prototypical random element \(X_i\), with mean \(\mu_i = \mathbb{E} X_i \in \mathcal{H}\) and covariance
operator \(\Sigma_i = \mathbb{E}\bigl[(X_i-\mu_i)\otimes(X_i-\mu_i)\bigr] : \mathcal{H}\to\mathcal{H}\), where \(u \otimes v\) denotes the operator \(w \mapsto \langle v, w\rangle\, u\); we observe \(n_i\) realizations from each population. In applications one inevitably works with finite-dimensional
representations of the mean and the covariance functions, which can be seen as functional analogues of the mean vector and variance-covariance matrix in multivariate data analysis. For the rest of the paper, when referring to the mean or the covariance operators, we will actually implicitly refer to their sample version, defined by

\begin{align*}
\widehat\mu_i &= \dfrac{1}{n_i} \sum_{j=1}^{n_i} X_{i,j}\\
\widehat\Sigma_i &= \dfrac{1}{n_i-1} 
\sum_{j=1}^{n_i} \bigl(X_{i,j}-\widehat\mu_i\bigr)\otimes\bigl(X_{i,j}-\widehat\mu_i\bigr).
\end{align*}

We refer to \citet{masarotto2019} for a justification of why the functional results employed in this paper hold for the finite-dimensional projections above.
The package proposes distance-based tools for inference which exploit
elements of optimal transport, in particular the so-called optimal
transport maps. Optimal transport is a mathematical field that studies how to optimally deform one distribution of mass to another, while minimizing the cost of such deformation. The optimal deformation is obtained through optimal transport maps. In the following we introduce only the few concepts of optimal transport needed by the package; see \citet{panaretos2020} for a book-length treatment of optimal transport methods in statistics.

A key definition is the
squared 2-Wasserstein distance between two Borel probability measures \(\mu\) and \(\nu\) on \(\mathcal H\). Let \(\Gamma(\mu,\nu)\) be the set of couplings of \(\mu\) and \(\nu\). These are Borel probability measures \(\pi\) on \(\mathcal H\times\mathcal H\) representing the collection of all possible joint probability setups with matching marginals \(\mu\) and \(\nu\). The squared 2-Wasserstein distance between \(\mu\) and \(\nu\) is defined as
\[
W_2^2(\mu,\nu)
=\inf_{\pi\in \Gamma(\mu,\nu)}
\int_{\mathcal H\times\mathcal H}{\|x-y\|^2}d{\pi(x,y)}.
\]
The distance is finite when \(\mu\) and \(\nu\) have a finite second moment. This optimization problem is known as the Monge--Kantorovich problem of optimal transportation, and admits a natural probabilistic formulation. Namely, if \(X\) and \(Y\) are random elements on \(\mathcal H\) with respective probability laws \(\mu\) and \(\nu\), then the problem translates to the minimization problem
\[
\inf_{Z_1\stackrel{d}{=}X,\,Z_2\stackrel{d}{=}Y}\mathbb E\|Z_1 - Z_2\|^2
\]
where the infimum is over all random vectors \((Z_1,Z_2)\) in \(\mathcal H\times \mathcal H\) such that \(X\stackrel{d}{=}Z_1\) and \(Y\stackrel{d}{=}Z_2\), marginally.

The infimum in \(W_2^2(\mu,\nu)\) does not have a closed form solution in general. A notable exception is in the Gaussian case. The 2-Wasserstein distance between centered Gaussian measures
\(\mu\sim N(0, \Sigma_1)\) and \(\nu \sim N(0, \Sigma_2)\) has the closed form \citep{dowson1982, olkin1982}:
\begin{equation}
W_2^2(\Sigma_1, \Sigma_2)=\inf_{\begin{smallmatrix} Z_1\sim N(0,\Sigma_1)\\Z_2 \sim N(0, \Sigma_2)\end{smallmatrix}} \mathbb{E}\|Z_1-Z_2\|^2 
 = \operatorname{tr}\Sigma_1 + \operatorname{tr}\Sigma_2
   - 2\operatorname{tr}\left[\left(\Sigma_2^{1/2} \Sigma_1
     \Sigma_2^{1/2}\right)^{1/2}\right],
 \label{eq:dist}
\end{equation}
where \(\Sigma_1\) and \(\Sigma_2\) are two covariance operators, or, after discretization on a common grid of \(M\)
points, two \(M \times M\) symmetric positive semi-definite matrices. The formula holds both in finite and infinite dimension.
Expression \eqref{eq:dist}
coincides with the squared Bures metric in quantum information theory and the squared Procrustes distance in shape analysis \citep{masarotto2019}. In the infinite-dimensional functional data setting, the formula extends by replacing matrix traces with operator traces.

Our testing procedure relies on the notion of Fréchet mean. An operator \({\overline\Sigma}\) is a Fréchet mean of the operators \(\{\Sigma_i\}_{i=1}^{K}\) if
\[\overline\Sigma=\underset{\Gamma\in G}{\arg\min}\sum_{i=1}^{K}W_2^2(\Sigma_i,\Gamma),\]
where \(G\) is the set of self-adjoint, non-negative, trace-class operators on \(\mathcal{H}\) --- after discretization, the closed convex cone of \(M\times M\) symmetric positive semi-definite matrices.
Fréchet means generalize arithmetic means to general metric spaces and can be thought of as barycenters of the operators \(\{\Sigma_i\}_{i=1}^{K}\).
In practice, we are required to compute an empirical Fréchet mean \(\hat\Sigma\),
\begin{equation}
\hat\Sigma=\underset{\mathbb{R}^{q\times q}\ni\Gamma\succeq 0}{\arg\min}\sum_{i=1}^{K}W_2^2(\widehat{\Sigma}_i,\Gamma).
\label{eq:empiricalfrechet}
\end{equation}
The minimizer exists, and it is unique as soon as at least one of the \(\widehat\Sigma_i\) is injective \citep{agueh2011, masarotto2019}; the rank-deficient case, ubiquitous with functional data, is discussed in the next section.

Further, the functional ANOVA test statistic relies on \emph{optimal transport maps}. Optimal maps are the workhorse of everything that
follows: they measure not only \emph{how far} two covariances are, but \emph{how}
one must be deformed into the other. The optimal map that deforms the Fréchet mean \({\overline\Sigma}\) into the operator \(\Sigma_j\) is given by:
\begin{equation}
    \mathbf{t}_j={\overline\Sigma}^{-1/2}({\overline\Sigma}^{1/2}\Sigma_j{\overline\Sigma}^{1/2})^{1/2}{\overline\Sigma}^{-1/2},\quad j=1,\ldots,K.
\end{equation}
In practice, the package employs the empirical versions of the \(\mathbf{t}_j\); when \(\overline\Sigma\) is singular, inverse square roots are understood on its range, as detailed in the next section. In \citet{masarotto2024}, the transport-map-based test was shown to compare favorably with existing second-order procedures, and to be particularly sensitive to high-frequency differences between covariances.

Optimal maps are also part of the computation of Fréchet means. In general, equation \eqref{eq:empiricalfrechet} has no closed-form solution, but its solution is characterized by the fixed-point property that the
weighted average of the optimal maps from \(\overline{\Sigma}\) to the
\(\Sigma_k\) is the identity,
\(\sum_k w_k\, \mathbf{t}_{\overline{\Sigma}}^{\Sigma_k} = I\). \citet{zemel2019} and \citet{masarotto2019}
turn this characterization into a steepest-descent iteration, provably convergent when at least one of the \(\Sigma_k\) (and the initial point) is injective,
\begin{equation}
\Sigma^{(j+1)} = \overline{T}^{(j)}\, \Sigma^{(j)}\, \overline{T}^{(j)},
\qquad
\overline{T}^{(j)} = \sum_{k} w_k\, \mathbf{t}_{\Sigma^{(j)}}^{\Sigma_k},
 \label{eq:iter}
\end{equation}
which is the algorithm implemented in the package.

Finally, the barycenter \({\overline\Sigma}\) provides an anchor at which the nonlinear geometry
can be linearized. The \emph{tangent space} at \(\overline{\Sigma}\) is a local linear approximation to the space of operators, and, for the sake of the package, can be thought as a canonical vector space. Mathematically, it contains the
log-images
\(\log_{\overline{\Sigma}}(\Sigma_k) = \mathbf{t}_{\overline{\Sigma}}^{\Sigma_k} - I\),
which are symmetric operators living in a genuine linear space; the
inverse (exponential) map is
\(\exp_{\overline{\Sigma}}(V) = (V + I)\,\overline{\Sigma}\,(V + I)\).
Ordinary linear methods --- averaging, PCA --- can be applied to the
log-images and their results pulled back to the space of covariances. This
is precisely what tangent space PCA does \citep{masarotto2024}.

\section{Package overview}\label{package-overview}

The fdWasserstein package (version 1.0, released 2024-02-06) is available from CRAN and can be installed and loaded with:

\begin{verbatim}
install.packages("fdWasserstein")
library("fdWasserstein")
\end{verbatim}

The package has no mandatory compiled dependencies and runs on any platform supporting R \citep{rcore} (\textgreater= 3.5.0). Optional parallelization uses the \CRANpkg{future} package \citep{bengtsson2021}. It exports five main functions
and one auxiliary one, summarized in Table \ref{tab:functions}, and ships
the phoneme dataset used throughout the next section.

\begin{table}[htbp]
\centering\small
\begin{tabular}{@{}>{\raggedright\arraybackslash}p{0.29\textwidth}>{\raggedright\arraybackslash}p{0.47\textwidth}>{\raggedright\arraybackslash}p{0.14\textwidth}@{}}
\toprule
Function & Description & Reference\\
\midrule
\texttt{dwasserstein} & Squared 2-Wasserstein distance between two covariance matrices & MPZ (2019)\\
\texttt{gaussBary} & Wasserstein barycenter (Fréchet mean) of $K$ covariance operators & MPZ (2019)\\
\texttt{wassersteinTest} & $K$-sample permutation or bootstrap test for equality of covariances & MPZ (2024)\\
\texttt{tangentPCA} & Tangent-space PCA of $K$ covariances at their barycenter & MPZ (2024)\\
\texttt{wassersteinCluster} & Soft clustering of curves by covariance structure & MM (2024)\\
\texttt{trimmedAverageSilhouette} & Silhouette-type index for selecting the number of clusters & MM (2024)\\
\bottomrule
\end{tabular}
\caption{\label{tab:functions}Summary of exported functions in \pkg{fdWasserstein}. MPZ = Masarotto, Panaretos and Zemel; MM = Masarotto and Masarotto.}
\end{table}

We gathered below some remarks on the implementation so that the case study
can be read fluently.

\textbf{Distance and rank deficiency.} \texttt{dwasserstein(A,\ B)} returns the
\emph{squared} distance \eqref{eq:dist}; users wanting the distance itself
should take a square root. All matrix square roots and inverse square
roots in the package are computed by spectral decomposition, discarding
eigenvalues below \texttt{sqrt(.Machine\$double.eps)}. Consequently every function
accepts rank-deficient inputs, so as not to fail when, as is often the case, covariances are estimated from fewer curves than grid points, as in the
example below where \(256 \times 256\) covariances are estimated from as few
as 40 curves. One caveat deserves to be stated explicitly: the uniqueness and convergence guarantees of Section 2 are established for injective operators, while the spectral truncation is what keeps all computations well defined --- effectively restricting them to the span of the retained eigenvectors --- in the singular case. In our experience the iteration \eqref{eq:iter}, started at the square-root mean described below, behaves well on rank-deficient inputs, as the examples of Section 4 illustrate; users should nonetheless be aware that this regime is not covered by the available theory.

\textbf{Barycenter.} \texttt{gaussBary(sigma,\ w)} takes an \(M \times M \times K\) array
and optional weights and runs iteration \eqref{eq:iter}, stopping when the
maximal absolute change falls below \texttt{eps} or after \texttt{max.iter} iterations.
The iteration is started, by default, at the Fréchet mean with respect to
the square-root distance, \(\left(\sum_k w_k \Sigma_k^{1/2}\right)^2\),
which is cheap, already respects positive semi-definiteness, and is
typically an excellent warm start; the call \texttt{gaussBary(sigma,\ max.iter\ =\ 0)}
returns it directly. Each iteration costs \(K\) spectral decompositions of
\(M \times M\) matrices, i.e., \(O(K M^3)\) time.

\textbf{Testing.} Given an \(n \times M\) matrix of discretized curves and a
vector of group labels, \texttt{wassersteinTest} tests the null hypothesis
\begin{equation}
\Sigma_1 = \cdots = \Sigma_K
\end{equation}
versus the alternative that at least one of the operators is different. The test rests on the fact that two covariance operators coincide exactly when the optimal map deforming the one centered Gaussian process into the other is the identity --- no deformation is needed --- so departures from the null can be measured by how far the fitted optimal maps are from the identity. Note that Gaussianity here is a device used to construct the statistic, \textbf{not} an assumption on the data. Groups are mean-centered separately
(\texttt{align\ =\ TRUE}), their sample covariances and weighted barycenter are
computed, and the default statistic contrasts the optimal maps with the
identity,
\begin{equation}
T = \sum_{k=1}^{K} n_k \left\|
\mathbf{t}_{\overline{\Sigma}}^{\widehat{\Sigma}_k} - I
\right\|^2_r,
 \label{eq:stat}
\end{equation}
with the Hilbert--Schmidt norm by default (argument \texttt{r}); the simpler
statistic \(\sum_k n_k\, d^2(\widehat{\Sigma}_k, \overline{\Sigma})\) is
available via \texttt{statistic\ =\ "distance"}. Statistic \eqref{eq:stat} is the
transport-based analogue of the classical ANOVA decomposition and
was found by \citet{masarotto2024} to be the most powerful among those compared. Moreover, it was found to be more powerful than its distance-based
counterpart, essentially because maps are sensitive to \emph{where} on the
spectrum two operators differ. The null distribution is obtained by
permuting (or, with \texttt{type\ =\ "bootstrap"}, resampling) the group labels of
the curves; because the permutation is applied after within-group centering, exchangeability under the null holds asymptotically rather than exactly \citep[see][ for the formal justification and finite-sample studies]{masarotto2024}. \texttt{B} statistic evaluations are performed, each requiring a
barycenter, so the cost is roughly \texttt{B} times that of a single fit. The
inner barycenter runs a fixed small number of iterations (\texttt{iter.bary},
default 10), which is ample for a test statistic and keeps the cost
predictable. Setting \texttt{use.future\ =\ TRUE} distributes the replicates over
any \pkg{future} backend.

\textbf{Tangent PCA.} When the hypothesis of equality of covariances is rejected, a natural question is to investigate the differences. Principal component analysis is a suitable tool to do that. \texttt{tangentPCA(sigma)} computes the barycenter, lifts each
covariance to the tangent space to obtain tangent space vectorized objects
\(V_k\), rescales by
\(\overline{\Sigma}^{1/2}\) so that the Euclidean inner product of the
vectorized lifts matches the Riemannian metric of the tangent space. After this, ordinary linear PCA can be carried out, before retracting the results to the space of covariance operators. Following \citet{masarotto2024}, the steps of the algorithm are as follows:

\begin{enumerate}
\item Compute the (empirical) Fréchet mean using `gaussBary(sigma, w)`, i.e.,
\[
\overline\Sigma
=\underset{\Gamma\succeq 0}{\arg\min} \sum_{j=1}^K W_2^2(\Gamma,\Sigma_j).
\]
\item Use the log map to lift
  $\Sigma_1,\dots,\Sigma_K$ to their log-images
$$V_j=\log_{\bar\Sigma}(\Sigma_j)=\mathbf{t}_{\overline\Sigma}^{\Sigma_j} - I, \qquad j = 1,\dots,K,$$
on the
  tangent space at the Fréchet mean, with $I$ being the identity operator. 
\item Perform linear PCA of the (suitably rescaled) tangent vectors $V_1,\dots,V_K$, obtaining principal directions $U_1, U_2, \dots$ in the tangent space.
\item Retract the principal directions onto the space of covariances via the exponential map,
  \[\exp_{\bar\Sigma}(U_m)=(U_m+I)\, \bar\Sigma\, (U_m +I).\]
\end{enumerate}

The implementation delegates the actual principal component computation to the built-in R function \texttt{prcomp}. The returned object is a genuine \texttt{prcomp} object
--- \texttt{summary}, \texttt{plot}, and the \texttt{\$x} scores behave exactly as R users
expect. The retraction of the principal components onto the manifold will give
principal geodesics that describe the main directions of variation of
the data on the manifold. For this reason, the values returned by \texttt{prcomp} are augmented with an array \texttt{\$eigenf} containing the leading
eigenvectors pulled back to covariance space via the exponential map, so
that principal modes of variation can be visualized as covariances rather
than as abstract tangent vectors. In the next Section, we provide a visualization of
the variation along the principal geodesics for the phoneme dataset.

\textbf{Soft clustering.} \texttt{wassersteinCluster(data,\ grp)} addresses a
different data structure: \(N\) \emph{samples} of curves (sample \(i\) containing
\(n_i\) curves, labeled by \texttt{grp}), each summarized by its sample covariance
\(\widehat{\Sigma}_i\) with \(\nu_i = n_i - 1\) degrees of freedom. For each
number of clusters \(k\) between \texttt{kmin} and \texttt{kmax} the procedure seeks to cluster the operators in a ``soft'' way - that is, allowing each operator to belong to more than one group. Soft clustering is particularly meaningful in cases in which the cluster separation is not neat and a soft classification, which allows for overlapping clusters, might be appropriate. Furthermore, it makes it easy to identify the groups that are most confused with each other.
In order to implement the clustering algorithm, we first need \(k\) prototypes of cluster barycenters \(\Gamma_1, \dots, \Gamma_k\) and a membership matrix
\(W = (w_{ij})\), \(w_{ij} \ge 0\), \(\sum_j w_{ij} = 1\), minimizing
\begin{equation}
\sum_{i=1}^{N} \nu_i \sum_{j=1}^{k} w_{ij}\,
d^2(\widehat{\Sigma}_i, \Gamma_j)
\qquad \text{subject to} \qquad
-\frac{1}{N}\sum_{i,j} w_{ij} \log w_{ij} = E,
 \label{eq:clust}
\end{equation}
i.e., a \(k\)-means-type criterion in the Wasserstein geometry in which hard
assignments are replaced by a partition matrix of prescribed average
entropy \(E\) \citep{masarotto2024clustering}. The entropy constraint is what makes the
clustering \emph{soft}: covariances lying between clusters can hedge their
membership, and the solution degrades gracefully when clusters overlap.
Internally, the constraint is enforced through its Lagrangian form: given
current distances, the optimal weights are
\(w_{ij} \propto \exp\{-\nu_i d^2(\widehat{\Sigma}_i, \Gamma_j)/\eta\}\),
with the temperature \(\eta\) found by \texttt{uniroot} so that the entropy
constraint is met. The criterion is minimized by block descent alternating
one fixed-point step \eqref{eq:iter} on each barycenter with a weight
update. Initial centers are chosen by a \(k\)-means\texttt{++}-type strategy
(sampling proportional to squared distance from the current centers),
improved by \texttt{nrefine} rounds of local swaps, and the whole search is
restarted \texttt{nstart} times; all pairwise distances between the
\(\widehat{\Sigma}_i\) are cached, so the initialization search costs at
most \(N(N-1)/2\) distance evaluations regardless of how many candidate
swaps are examined.

The entropy level \(E\) and the search parameters \texttt{nstart} and \texttt{nrefine} are chosen by the user. The defaults, suggested by the numerical experiments in \citet{masarotto2024clustering}, are \texttt{nstart\ =\ 5}, \texttt{nrefine\ =\ 5}, and \(E = -0.75\,(0.95 \log 0.95 + 0.05 \log 0.05) + 0.25 \log 2 \approx 0.32\): the average entropy obtained when three quarters of the units are assigned nearly hard, with weights \((0.95, 0.05)\) on their two closest clusters, and the remaining quarter is split evenly between two clusters.

The package includes also an easy adaptation of
the methodology that supports fast clustering when the number of covariance
operators is very large: \texttt{nreduced} allows the above search to
be conducted on a random subset of the samples, with all memberships
recomputed at the end.

Finally, the package also offers guidance on choosing the most
suitable number of clusters. The helper \texttt{trimmedAverageSilhouette} computes, for
each \(k\), a silhouette-type index \(s_i = 1 - d_{(1)}^2(i) / d_{(2)}^2(i)\)
--- where \(d_{(1)}, d_{(2)}\) are the distances from \(\widehat{\Sigma}_i\)
to the two nearest barycenters --- averaged over the samples whose maximal
membership weight is above average (the ``trimming''); the number of
clusters maximizing this TASW index is retained.

It is also possible to assess whether any cluster structure is present at all, with the optional permutation
test (\texttt{nperm}).

\section{Case study: the phoneme data}\label{case-study-the-phoneme-data}

\begin{figure}

{\centering \includegraphics[width=1\linewidth]{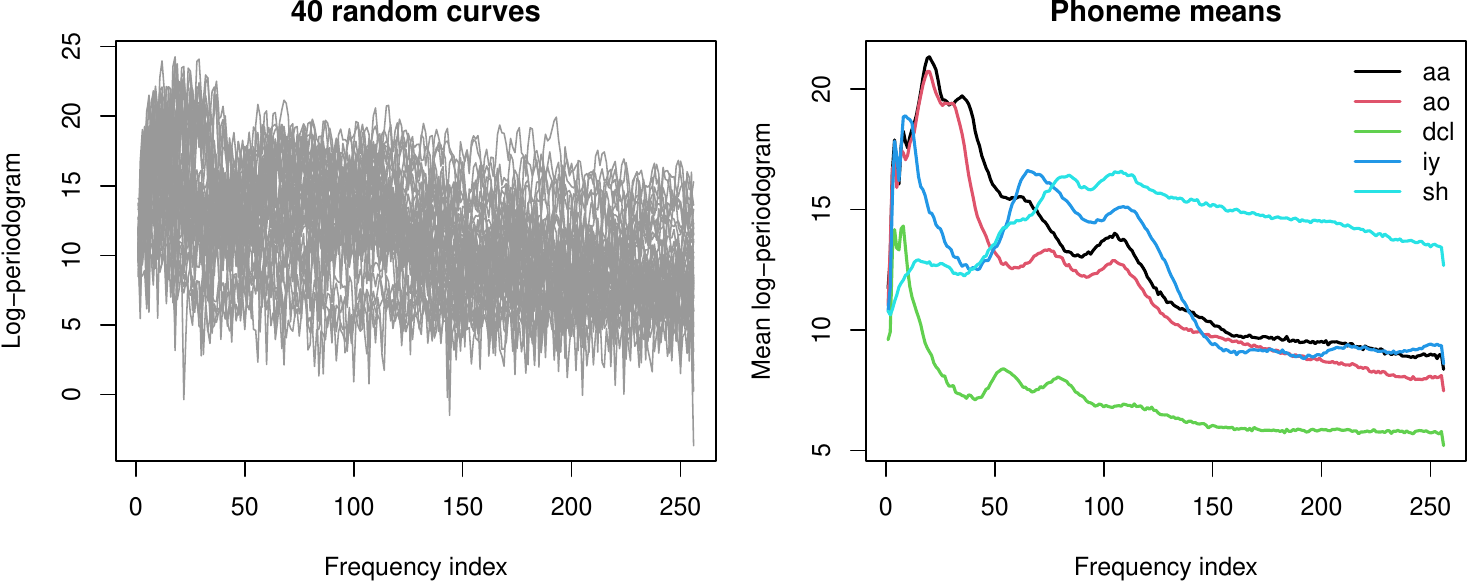} 

}

\caption{Left: 40 randomly chosen log-periodograms. Right: mean log-periodogram of each phoneme.}\label{fig:datafig}
\end{figure}

The phoneme data, extracted from the TIMIT corpus and popularized by
\citet{hastie1995}, consist of 4509 log-periodograms of length \(M = 256\),
computed from 32 ms recordings of male speakers pronouncing one of five
phonemes: the vowels /aa/ (as in ``dark'', 695 curves) and /ao/ (as in
``water'', 1022), the stop consonant /dcl/ (as in ``dark'', 757), the vowel
/iy/ (as in ``she'', 1163), and the fricative /sh/ (as in ``she'', 872).
Loading the data provides the \(4509 \times 256\) matrix \texttt{logPeriodogram}
and the label vector \texttt{Phoneme}:

\begin{verbatim}
library(fdWasserstein)
data(phoneme)
table(Phoneme)
\end{verbatim}

\begin{verbatim}
Phoneme
  aa   ao  dcl   iy   sh 
 695 1022  757 1163  872 
\end{verbatim}

Figure \ref{fig:datafig} displays a random subset of the curves and the
five mean log-periodograms. The means differ visibly, but our interest
here is in the second-order structure: how the curves \emph{fluctuate} around
their group means. This is the natural question for this kind of data ---
the timbre of a phoneme is encoded in the covariation of energy across
frequencies as much as in the mean spectrum --- and it is the question the
package is designed to answer.

\subsection{Distances and barycenter}\label{distances-and-barycenter}

We first compute the five within-phoneme sample covariances and their
pairwise squared distances \eqref{eq:dist}. At the level of our sample, we have access to empirical versions of \(\{\widehat{\Sigma}_j\}\), constructed from the \(n_i\) curves of each group. These could simply be the empirical covariances within each group (under a complete observation assumption), or some smoothed estimator (for instance the empirical covariance of smoothed versions of the \(\{X_{ij}\}\) \citep{ramsay2005}, or PACE-type estimators \citep{yao2005functional}). Whichever the case may be, the \(\widehat{\Sigma}_j\) are finite dimensional, of rank \(q\leq n_i\). In case a smoothing technique is used, we assume that it is such that the \(\widehat{\Sigma}_j\) share a common range, and can thus be represented as \(q\times q\) positive matrices, via a common (tensor product) basis. For tidiness, we use the same notation for \(\widehat{\Sigma}_j\) and its \(q\times q\) matrix representation in the common basis.
If the \(\widehat{\Sigma}_j\) are obtained from smoothed versions of the \(\{X_{ij}\}\), the operators can be computed with the command \texttt{var.fd} in the \CRANpkg{fda} package \citep{fda-pkg}. In our case, we compute the covariances directly from the data.

\begin{verbatim}
gg <- sort(unique(Phoneme))
Sig <- sapply(gg, function(l) cov(logPeriodogram[Phoneme == l, ]),
              simplify = "array")
D <- matrix(0, 5, 5, dimnames = list(gg, gg))
for (i in 1:4) for (j in (i+1):5)
    D[i, j] <- D[j, i] <- dwasserstein(Sig[, , i], Sig[, , j])
round(D, 1)
\end{verbatim}

\begin{verbatim}
       aa    ao   dcl    iy    sh
aa    0.0 119.1 286.2 156.4 340.8
ao  119.1   0.0 245.2 166.4 287.3
dcl 286.2 245.2   0.0 304.8 187.3
iy  156.4 166.4 304.8   0.0 322.9
sh  340.8 287.3 187.3 322.9   0.0
\end{verbatim}

The computation takes about a second. The table is phonetically sensible:
the two back vowels /aa/ and /ao/, which are hard to
distinguish, are the closest pair, while the fricative /sh/ is far
from all vowels. The visual plots of the five covariance operators appear below.

\begin{verbatim}
zl <- range(Sig)
op <- par(mfrow = c(1, 5), mar = c(2, 2, 2, 1))
invisible(Map(function(i, nm)
  image(Sig[, , i], main = nm, col = topo.colors(64), zlim = zl, axes = FALSE),
  seq_len(dim(Sig)[3]), dimnames(Sig)[[3]]))
\end{verbatim}

\begin{figure}

{\centering \includegraphics[width=1\linewidth]{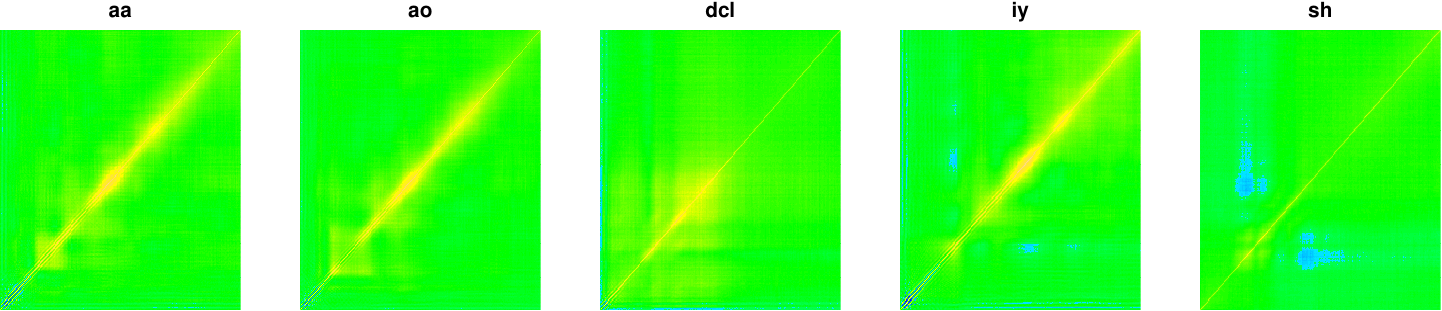} 

}

\caption{Covariance operators of the five phonemes (shared color scale).}\label{fig:cov-image}
\end{figure}

\begin{verbatim}
par(op)
\end{verbatim}

The function \texttt{gaussBary} computes the Fréchet mean of the five covariances, weighted by
their degrees of freedom:

It converges in a handful of iterations:

\begin{verbatim}
bary$iter
\end{verbatim}

\begin{verbatim}
[1] 8
\end{verbatim}

Each iteration performs five spectral decompositions of \(256 \times 256\)
matrices; the whole computation is a little shorter than 2 seconds on one core of a
standard laptop-class machine (the setting for all timings quoted below).

\subsection{Testing equality of the covariance operators}\label{testing-equality-of-the-covariance-operators}

Is the second-order structure really different across phonemes? The
ANOVA-type test with the map-based statistic \eqref{eq:stat} is invoked
directly on the curves:

\begin{verbatim}
set.seed(1001)
glob <- wassersteinTest(logPeriodogram, Phoneme, B = 100, r = "HS")
c(statistic = glob$stat, p.value = glob$p.value)
\end{verbatim}

\begin{verbatim}
   statistic      p.value 
9.955038e+04 9.900990e-03 
\end{verbatim}

The observed statistic, 99550, is nowhere
near the permutation distribution, whose 100 replicates range
from 61390 to
63010; the p-value
\((1 + \#\{T^\ast \ge T\})/(1 + B) = 1/101\) is the smallest attainable with
\(B = 100\). Each statistic evaluation requires the five sample covariances,
their barycenter, and the five optimal maps, about 3 seconds here, so the
full test runs in about five minutes on a single core;
\texttt{use.future\ =\ TRUE} parallelizes the permutations. We use \(B = 100\) so that the article reproduces quickly; for reporting purposes a larger number of replicates, say \(B = 999\), is preferable and remains entirely practical with parallelization.

A more delicate question is whether the two closest phonemes can be told
apart by their covariances alone. Restricting the test to /aa/ and /ao/:

\begin{verbatim}
sel <- Phoneme %in% c("aa", "ao")
set.seed(1001)
pw <- wassersteinTest(logPeriodogram[sel, ], Phoneme[sel],
                      B = 100, r = "HS")
c(statistic = pw$stat, p.value = pw$p.value)
\end{verbatim}

\begin{verbatim}
   statistic      p.value 
1.795118e+04 9.900990e-03 
\end{verbatim}

We can see again a decisive rejection (permutation replicates range from
14989 to
15842). In conclusion, we can detect a marked difference in the covariance
structure, even in the ``difficult'' pair that mean-based classifiers
struggle to differentiate.

\subsection{Tangent space principal component analysis}\label{tangent-space-principal-component-analysis}

If the test of equality is rejected, a natural question is to investigate the variability of the sample of covariances around its
Fréchet mean, and possibly to interpret the main directions of this variation. As in multivariate analysis in Euclidean spaces, Principal Component Analysis
(PCA) is a good candidate for such tasks. For functional data, this has been extended
to functional PCA \citep{ramsay2005}.
One way of carrying out fPCA in Hilbert spaces is based on the eigenstructure of the covariance
operator, by analogy to PCA in finite dimensions. When the space is non-Euclidean, as in the case of a sample of covariances,
one way to carry out PCA is by lifting the analysis onto the tangent space. In finite dimensions,
this is known as tangent space PCA \citep[see e.g.][]{dryden2009}. The tangent space provides a local linear approximation to the curved space of covariances.
Once the covariances are mapped onto the tangent space through the log map,
they can be uniquely identified with a tangent vector that belongs to a linear space, and
therefore, a standard linear (functional) PCA can be carried out.

In order to test our methodology in a more realistic situation, we artificially enlarge our sample of covariances by subsampling the original data. Specifically, from each of the five phonemes we repeatedly draw 50 log-periodogram curves at random and form the sample covariance of the draw. Twelve independent draws per phoneme yield 60 covariance operators of
size \(256 \times 256\) (each of rank at most 49). Since the draws from a given phoneme overlap, the resulting covariances are positively dependent within phonemes; this is immaterial for the illustration, but worth keeping in mind when reading the figures. We then carry out the PCA on these 60
covariances:

\begin{verbatim}
set.seed(12345)
nsub <- 12; n <- 50
gg <- unique(Phoneme)   # sh, iy, dcl, aa, ao
M <- NCOL(logPeriodogram)
Sigma <- array(dim = c(M, M, length(gg) * nsub))
r <- 0
for (l in gg) for (i in 1:nsub) {
    r <- r + 1
    Sigma[, , r] <- cov(logPeriodogram[sample(which(Phoneme == l), n), ])
}
pca <- tangentPCA(Sigma, max.iter = 3)
round(100 * pca$sdev[1:4]^2 / sum(pca$sdev^2), 1)
\end{verbatim}

\begin{verbatim}
[1] 9.9 5.7 4.1 3.1
\end{verbatim}

\begin{figure}

{\centering \includegraphics[width=1\linewidth]{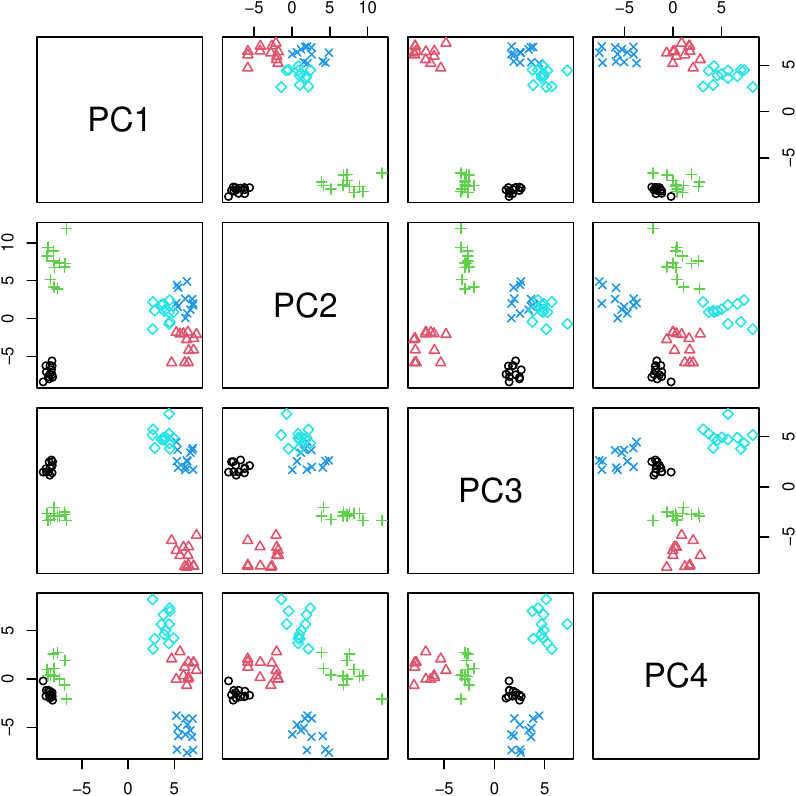} 

}

\caption{Scores of the 60 subsampled phoneme covariances on the first four tangent principal components. Colors and symbols denote the phonemes, in the order /sh/ (black circles), /iy/ (red triangles), /dcl/ (green plus), /aa/ (blue crosses), /ao/ (cyan diamonds).}\label{fig:pcafig}
\end{figure}

The computation takes about half a minute, dominated by the barycenter and
the 60 liftings; \texttt{max.iter\ =\ 3} suffices here because the square-root warm start of Section 3 already lands close to the barycenter. Since \texttt{pca} is a \texttt{prcomp} object, the standard idioms
apply; the first four components explain the percentages of tangent-space
variance printed above. The percentages are modest --- the leading component accounts for about a tenth of the total tangent-space variance --- because the denominator is dominated by subsampling noise spread over the many remaining directions; the between-phoneme signal is nevertheless concentrated in the first few components, as the score plots make plain. Figure \ref{fig:pcafig} displays the scores on
the first four components, with symbols identifying the phonemes ---
information the PCA itself never saw. The five phonemes form five clearly
separated clusters already in the first two or three components. We can see that each phoneme is isolated in
at least one plot, with /aa/ and
/ao/ sitting closest, in agreement with the distance table. The array
\texttt{pca\$eigenf} contains the corresponding principal modes of variation as
bona fide covariances, obtained by the exponential map, which can be
plotted as images or, as in \citet{masarotto2024}, summarized through their
leading eigenfunctions.

\subsection{Soft clustering of the samples}\label{soft-clustering-of-the-samples}

Finally we reverse the point of view: suppose the phoneme labels were
unknown, and we were handed \(N = 75\) \emph{samples} of curves --- 15 samples
per phoneme, each containing \(n = 40\) curves. Can the phonemes be
recovered from the covariance structure alone? This is the setting of
\texttt{wassersteinCluster}, which receives the raw curves and the sample
identifiers. The code below illustrates the Trimmed Average Silhouette Width (TASW) index of \citet{masarotto2024clustering}, used to estimate the number of clusters; the highest value indicates the most plausible number of clusters.

\begin{verbatim}
set.seed(12345)
nsub <- 15; n <- 40
N <- n * nsub * length(gg)
X <- matrix(NA, N, M); gr <- integer(N)
r <- 1; first <- 1; last <- n
for (l in gg) for (i in 1:nsub) {
    X[first:last, ] <- logPeriodogram[sample(which(Phoneme == l), n), ]
    gr[first:last] <- r
    r <- r + 1; first <- first + n; last <- last + n
}
a <- wassersteinCluster(X, gr, kmin = 2, kmax = 8, verbose = FALSE)
round(trimmedAverageSilhouette(a, plot = FALSE), 3)
\end{verbatim}

\begin{verbatim}
    2     3     4     5     6     7     8 
0.284 0.350 0.342 0.390 0.369 0.374 0.383 
\end{verbatim}

\begin{figure}

{\centering \includegraphics[width=1\linewidth]{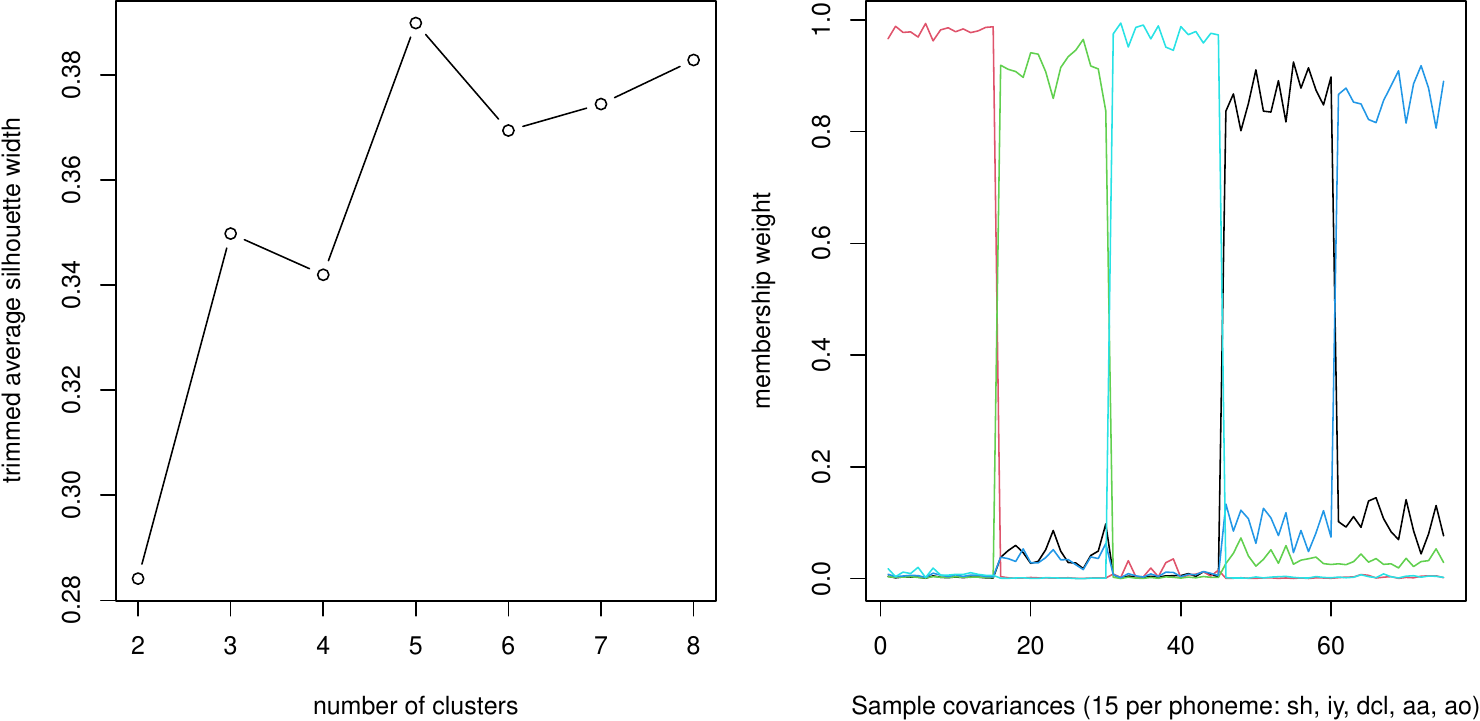} 

}

\caption{Left: trimmed average silhouette width as a function of the number of clusters; the maximum selects $k = 5$. Right: membership weights of the $k = 5$ solution across the 75 sample covariances (15 per phoneme, in the order /sh/, /iy/, /dcl/, /aa/, /ao/).}\label{fig:clustfig}
\end{figure}

The complete run --- pairwise distances between the 75 sample covariances,
initialization and block descent for each \(k\) from 2 to 8 --- takes about
twenty minutes on a single core; each \(k\) converged in 7 or 8
block-descent iterations. The TASW criterion is maximized at \(k = 5\), the
true number of phonemes (Figure \ref{fig:clustfig}, left panel). The
\(k = 5\) solution is stored in \texttt{a{[}{[}4{]}{]}} (element \(k - \texttt{kmin} + 1 = 4\) of the returned list); its membership matrix reveals an
essentially perfect reconstruction:

\begin{verbatim}
w5 <- a[[4]]$w
truth <- rep(gg, each = nsub)
table(truth, hard = apply(w5, 1, which.max))
\end{verbatim}

\begin{verbatim}
     hard
truth  1  2  3  4  5
  aa  15  0  0  0  0
  ao   0  0  0 15  0
  dcl  0  0  0  0 15
  iy   0  0 15  0  0
  sh   0 15  0  0  0
\end{verbatim}

All 75 samples are assigned to the correct phoneme, with maximal
membership weights ranging from
0.8 to
0.99 (median
0.92) --- large, but deliberately
not equal to one: the entropy constraint keeps the assignment soft, and
the smallest weights occur precisely for the /aa/ and /ao/ samples, the
overlapping pair. Figure \ref{fig:clustfig} traces the five membership
profiles across the 75 samples, ordered by phoneme, and displays the
characteristic block structure of a well-separated soft partition.

\section{Summary}\label{summary}

\pkg{fdWasserstein} equips R users with a coherent set of tools for the
statistical analysis of covariance operators of functional data under the
Wasserstein--Procrustes geometry: distances and optimal maps, barycenters,
an ANOVA-type permutation test, tangent space PCA, and covariance-based
soft clustering, all operating directly on discretized curves or on arrays
of covariances, and all robust to the rank deficiency inherent in
functional data. On the phoneme data, the complete workflow --- test,
exploration, clustering --- runs in well under an hour on a single core,
scales further through the \pkg{future} framework, and recovers the known
structure of the data with striking fidelity.

Two directions for future development are natural. First, regularized
transport maps for severely ill-conditioned barycenters; second, methods
for covariance operators varying smoothly with a covariate, the regression
counterpart of the ANOVA implemented here.

\section{Acknowledgments}\label{acknowledgments}

The author would like to thank Y. Zemel for an earlier version of the gausbary algorithm, and is grateful to G. Masarotto for the advice throughout the development of the \pkg{fdWasserstein} package.

\bibliography{fdwasserstein.bib}

\address{%
Valentina Masarotto\\
Mathematical Institute, Leiden University\\%
Einsteinweg 55, 2333 CC Leiden\\ The Netherlands\\
\href{mailto:v.masarotto@math.leidenuniv.nl}{\nolinkurl{v.masarotto@math.leidenuniv.nl}}%
}

%% file: RJwrapper-arxiv.bbl
\begin{thebibliography}{24}
\providecommand{\natexlab}[1]{#1}
\providecommand{\url}[1]{\texttt{#1}}
\expandafter\ifx\csname urlstyle\endcsname\relax
  \providecommand{\doi}[1]{doi: #1}\else
  \providecommand{\doi}{doi: \begingroup \urlstyle{rm}\Url}\fi

\bibitem[Agueh and Carlier(2011)]{agueh2011}
M.~Agueh and G.~Carlier.
\newblock Barycenters in the {W}asserstein space.
\newblock \emph{SIAM Journal on Mathematical Analysis}, 43\penalty0
  (2):\penalty0 904--924, 2011.
\newblock \doi{10.1137/100805741}.

\bibitem[Bengtsson(2021)]{bengtsson2021}
H.~Bengtsson.
\newblock A unifying framework for parallel and distributed processing in {R}
  using futures.
\newblock \emph{The R Journal}, 13\penalty0 (2):\penalty0 208--227, 2021.
\newblock \doi{10.32614/RJ-2021-048}.

\bibitem[Bhatia et~al.(2019)Bhatia, Jain, and Lim]{bhatia2019}
R.~Bhatia, T.~Jain, and Y.~Lim.
\newblock On the {B}ures--{W}asserstein distance between positive definite
  matrices.
\newblock \emph{Expositiones Mathematicae}, 37\penalty0 (2):\penalty0 165--191,
  2019.
\newblock \doi{10.1016/j.exmath.2018.01.002}.

\bibitem[Chen et~al.(2024)]{frechet-pkg}
Y.~Chen et~al.
\newblock \emph{frechet: Statistical Analysis for Random Objects and
  Non-Euclidean Data}, 2024.
\newblock URL \url{https://CRAN.R-project.org/package=frechet}.
\newblock R package.

\bibitem[Dowson and Landau(1982)]{dowson1982}
D.~C. Dowson and B.~V. Landau.
\newblock The {F}r\'echet distance between multivariate normal distributions.
\newblock \emph{Journal of Multivariate Analysis}, 12\penalty0 (3):\penalty0
  450--455, 1982.
\newblock \doi{10.1016/0047-259X(82)90077-X}.

\bibitem[Dryden et~al.(2009)Dryden, Koloydenko, and Zhou]{dryden2009}
I.~L. Dryden, A.~Koloydenko, and D.~Zhou.
\newblock Non-{E}uclidean statistics for covariance matrices, with applications
  to diffusion tensor imaging.
\newblock \emph{The Annals of Applied Statistics}, 3\penalty0 (3):\penalty0
  1102--1123, 2009.
\newblock \doi{10.1214/09-AOAS249}.

\bibitem[G{\'o}recki and Smaga(2019)]{gorecki2019}
T.~G{\'o}recki and {\L}.~Smaga.
\newblock {fdANOVA}: An {R} software package for analysis of variance for
  univariate and multivariate functional data.
\newblock \emph{Computational Statistics}, 34:\penalty0 571--597, 2019.
\newblock \doi{10.1007/s00180-018-0842-7}.

\bibitem[Hastie et~al.(1995)Hastie, Buja, and Tibshirani]{hastie1995}
T.~Hastie, A.~Buja, and R.~Tibshirani.
\newblock Penalized discriminant analysis.
\newblock \emph{The Annals of Statistics}, 23\penalty0 (1):\penalty0 73--102,
  1995.
\newblock \doi{10.1214/aos/1176324456}.

\bibitem[Masarotto and Chen(2024)]{masarotto2024tonal}
V.~Masarotto and Y.~Chen.
\newblock Tonal coarticulation revisited: functional covariance analysis to
  investigate the planning of co-articulated tones by standard chinese
  speakers.
\newblock \emph{arXiv preprint arXiv:2409.01194}, 2024.

\bibitem[Masarotto and Masarotto(2024)]{masarotto2024clustering}
V.~Masarotto and G.~Masarotto.
\newblock Covariance-based soft clustering of functional data based on the
  {W}asserstein--{P}rocrustes metric.
\newblock \emph{Scandinavian Journal of Statistics}, 51\penalty0 (2), 2024.
\newblock \doi{10.1111/sjos.12692}.

\bibitem[Masarotto et~al.(2019)Masarotto, Panaretos, and Zemel]{masarotto2019}
V.~Masarotto, V.~M. Panaretos, and Y.~Zemel.
\newblock Procrustes metrics on covariance operators and optimal transportation
  of {G}aussian processes.
\newblock \emph{Sankhya A}, 81\penalty0 (1):\penalty0 172--213, 2019.
\newblock \doi{10.1007/s13171-018-0130-1}.

\bibitem[Masarotto et~al.(2024)Masarotto, Panaretos, and Zemel]{masarotto2024}
V.~Masarotto, V.~M. Panaretos, and Y.~Zemel.
\newblock Transportation-based functional {ANOVA} and {PCA} for covariance
  operators.
\newblock \emph{Electronic Journal of Statistics}, 18\penalty0 (1):\penalty0
  1887--1916, 2024.
\newblock \doi{10.1214/24-EJS2240}.

\bibitem[Olkin and Pukelsheim(1982)]{olkin1982}
I.~Olkin and F.~Pukelsheim.
\newblock The distance between two random vectors with given dispersion
  matrices.
\newblock \emph{Linear Algebra and its Applications}, 48:\penalty0 257--263,
  1982.
\newblock \doi{10.1016/0024-3795(82)90112-4}.

\bibitem[Panaretos and Zemel(2020)]{panaretos2020}
V.~M. Panaretos and Y.~Zemel.
\newblock \emph{An Invitation to Statistics in {W}asserstein Space}.
\newblock SpringerBriefs in Probability and Mathematical Statistics. Springer,
  Cham, 2020.
\newblock \doi{10.1007/978-3-030-38438-8}.

\bibitem[Panaretos et~al.(2010)Panaretos, Kraus, and Maddocks]{panaretos2010}
V.~M. Panaretos, D.~Kraus, and J.~H. Maddocks.
\newblock Second-order comparison of {G}aussian random functions and the
  geometry of {DNA} minicircles.
\newblock \emph{Journal of the American Statistical Association}, 105\penalty0
  (490):\penalty0 670--682, 2010.
\newblock \doi{10.1198/jasa.2010.tm09239}.

\bibitem[Pigoli et~al.(2014)Pigoli, Aston, Dryden, and Secchi]{pigoli2014}
D.~Pigoli, J.~A.~D. Aston, I.~L. Dryden, and P.~Secchi.
\newblock Distances and inference for covariance operators.
\newblock \emph{Biometrika}, 101\penalty0 (2):\penalty0 409--422, 2014.
\newblock \doi{10.1093/biomet/asu008}.

\bibitem[{R Core Team}(2024)]{rcore}
{R Core Team}.
\newblock \emph{R: A Language and Environment for Statistical Computing}.
\newblock R Foundation for Statistical Computing, Vienna, Austria, 2024.
\newblock URL \url{https://www.R-project.org/}.

\bibitem[Ramsay and Silverman(2005)]{ramsay2005}
J.~O. Ramsay and B.~W. Silverman.
\newblock \emph{Functional Data Analysis}.
\newblock Springer, New York, 2nd edition, 2005.

\bibitem[Ramsay et~al.(2024)Ramsay, Graves, and Hooker]{fda-pkg}
J.~O. Ramsay, S.~Graves, and G.~Hooker.
\newblock \emph{fda: Functional Data Analysis}, 2024.
\newblock URL \url{https://CRAN.R-project.org/package=fda}.
\newblock R package.

\bibitem[Schuhmacher et~al.(2024)]{transport-pkg}
D.~Schuhmacher et~al.
\newblock \emph{transport: Computation of Optimal Transport Plans and
  {W}asserstein Distances}, 2024.
\newblock URL \url{https://CRAN.R-project.org/package=transport}.
\newblock R package.

\bibitem[Wood(2017)]{wood2017}
S.~N. Wood.
\newblock \emph{Generalized Additive Models: An Introduction with {R}}.
\newblock Chapman and Hall/CRC, Boca Raton, 2nd edition, 2017.

\bibitem[Yao et~al.(2005)Yao, M{\"u}ller, and Wang]{yao2005functional}
F.~Yao, H.-G. M{\"u}ller, and J.-L. Wang.
\newblock Functional data analysis for sparse longitudinal data.
\newblock \emph{Journal of the American Statistical Association}, 100\penalty0
  (470):\penalty0 577--590, 2005.

\bibitem[You(2024)]{t4transport-pkg}
K.~You.
\newblock \emph{T4transport: Tools for Computational Optimal Transport}, 2024.
\newblock URL \url{https://CRAN.R-project.org/package=T4transport}.
\newblock R package.

\bibitem[Zemel and Panaretos(2019)]{zemel2019}
Y.~Zemel and V.~M. Panaretos.
\newblock Fr\'echet means and {P}rocrustes analysis in {W}asserstein space.
\newblock \emph{Bernoulli}, 25\penalty0 (2):\penalty0 932--976, 2019.
\newblock \doi{10.3150/17-BEJ1009}.

\end{thebibliography}
